\documentclass[a4paper,11pt]{revtex4}
\usepackage{pdfpages}
\usepackage{graphicx}
\usepackage{amsmath}
\usepackage{amsfonts}
\usepackage{amssymb}
\usepackage[colorlinks]{hyperref}
\hypersetup{linkcolor=black, citecolor=black, filecolor=black, urlcolor=black}
\usepackage[all]{hypcap}

\newcommand{\IOPP}{Key Laboratory of Quark $\&$ Lepton Physics (MOE) and Institute of Particle Physics, \\
Central China Normal University, Wuhan 430079, China}

\begin{document}

\title{Energy Dependence of Light Nuclei ($d$, $t$) Production at STAR}

\author{Dingwei \textsc{Zhang} (for the STAR Collaboration)}

\affiliation{\IOPP}
\email{zhangdingwei@mails.ccnu.edu.cn}

\begin{abstract}
In high-energy nuclear collisions, the production of light nuclei is sensitive to the temperature and phase-space density of the system at freeze-out. In addition, the phase transition from QGP to the hadronic phase will lead to large baryon density fluctuation, which will be reflected in the light nuclei production. For example, the ratio of proton ($N(p)$) and triton ($N(t)$) to deuteron ($N(d)$) yields, which is defined as $N(t)$$\cdot$$N(p)$/$N^2(d)$, may be used as a sensitive observable to search for the QCD critical point. In this paper, we will report the energy dependence of light nuclei ($d$, $t$) production in Au+Au collisions at $\sqrt{s_{NN}}$ =7.7, 11.5, 14.5, 19.6, 27, 39, 62.4, and 200 GeV measured by the STAR experiment at RHIC. We will present the beam energy dependence for the coalescence parameter $B_2(d)$ and $B_3(t)$, particle ratios ($d/p$, $t/p$, and $t/d$), and the yield ratio of $N(t)$$\cdot$$N(p)$/$N^2(d)$. Their physics implications will be discussed.
\end{abstract}

\maketitle
\section{Introduction}
In relativistic heavy-ion collisions, the most important goal is to study the properties of nuclear matter at high baryon density or temperature. The phase transition between the Quark-Gluon Plasma (QGP) and the hadronic matter is currently a topic of great interest~\cite{Fukushima:2013rx}. Light nuclei are sensitive small binding energy, they are sensitive to the local nucleon density. Light nuclei production is described by coalescence~\cite{Csernai:1986qf} and thermodynamic models~\cite{Mekjian:1978zz}.  In the coalescence picture, the density of the cluster is proportional to the proton density times the probability of finding a neutron within a small sphere of radius around the proton momentum. Nucleon coalescence mechanism can be described as: 

\begin{equation}
	{E_A}\frac{{{d^3}{N_A}}}{{{d^3}{p_A}}} = {B_A}{\left( {{E_p}\frac{{{d^3}{N_p}}}{{{d^3}{p_p}}}} \right)^Z}{\left( {{E_n}\frac{{{d^3}{N_n}}}{{{d^3}{p_n}}}} \right)^{A - Z}} \approx {B_A}{\left( {{E_p}\frac{{{d^3}{N_p}}}{{{d^3}{p_p}}}} \right)^A},
\end{equation}
where ${E_p}\frac{{{d^3}{N_p}}}{{{d^3}{p_p}}}$ is the Lorentz-invariant momentum distribution of proton, $A$ is the mass number, $Z$ is the proton number, $p_A = Ap_p$. The coalescence parameter, $B_A$, reflects the probability of nucleon coalescence, which is related to the local nucleon density~\cite{Adam:2019wnb}.

Based on the coalescence picture, the yield ratio, $N(t)$$\cdot$$N(p)$/$N^2(d)$, is sensitive to the neutron density fluctuation, $\Delta n$=$\langle(\delta n)^2\rangle/\langle n\rangle^2$, at kinetic freeze-out in relativistic heavy-ion collisions, where $\langle n\rangle$ denotes the average value over space and $\delta n$ denotes the fluctuation of neutron density from its average value $\langle n\rangle$~\cite{Sun:2018jhg}. In this case, the yield ratio of light nuclei can be approximated as: 
\begin{equation}
	N(t) \cdot N(p)/N^2(d)=g(1+\Delta n), with\ g = 0.29.
\end{equation}

In heavy-ion collisions, the created matter is expected to develop strong baryon density fluctuation~\cite{Stephanov:1998dy}. When its evolution trajectory in the QCD phase diagram passes across the first-order phase transition line,  a rapid increase of correlation length in the critical region has emerged as a results of the spinodal instability or approaches the CEP~\cite{Sun:2018jhg}. Therefore, light nuclei production at kinetic freeze-out in relativistic heavy-ion collisions may provide a unique probe to the critical endpoint in the QCD phase diagram.
\section{Results and discussions}
\subsection{Transverse momentum spectra}
\begin{figure}[htpb]
\centering
\includegraphics[width=1.\textwidth]{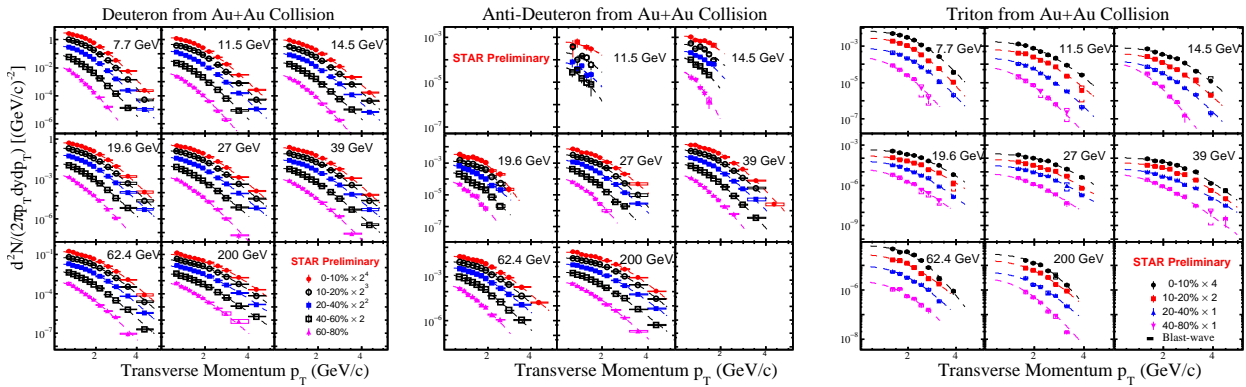}
\caption{\label{spectra}Transverse momentum spectra of $d$ (left), $\bar d$ (middle), and $t$ (right) measured at midrapidity in Au+Au collisions at $\sqrt{s_\mathrm{NN}}=$7.7-200 GeV for different collision centralities. The dashed lines are the individual fits to the data with blast-wave functions. Both the statistical and systematic error are shown.}
\end{figure}
The results are based on RHIC Beam Energy Scan I (BES-I) and are obtained for Au+Au collisions at $\sqrt{s_\mathrm{NN}}=$7.7-200 GeV at midrapidity. The main detectors used to obtain the results are the Time Projection Chamber (TPC)~\cite{Anderson:2003ur} and Time-Of-Flight (TOF) detectors~\cite{Llope:2005yw}. The transverse momentum spectra of identified $d$ (left panel), $\bar d$ (middle panel), and $t$ (right panel) are shown in Fig.~\ref{spectra}. Due to limited statistics, there are not enough candidates to show the $\bar d$ spectra at $\sqrt{s_{NN}}$  = 7.7 GeV~\cite{Adam:2019wnb}. The spectra are fitted with individual blast-wave functions and show a hardening with increasing multiplicity/centrality~\cite{Schnedermann:1993ws}. The blast-wave fits are shown as dashed lines in Fig.~\ref{spectra}. 
\subsection{Coalescence parameters}
\begin{figure}[!htp]
\centering
\includegraphics[width=0.78\textwidth]{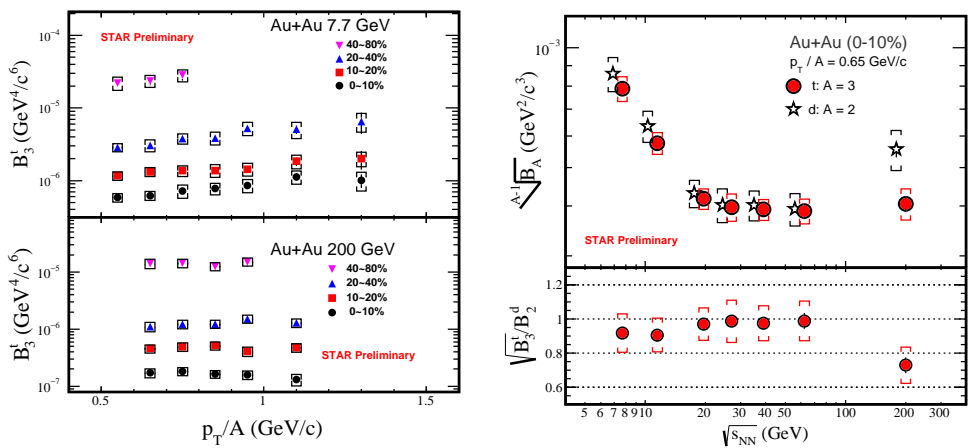}
\caption{\label{coal}(Left panel) Coalescence parameter $B_3$ as a function of $p_T/A$ for triton from 7.7 and 200 GeV in 0-10\%, 10-20\%, 20-40\% and 40-80\% central Au+Au collisions. (Right panel) Coalescence parameter $B_2$ (open star) and $\sqrt{B_3}$ (red dot) as a function of collision energy in 0-10\% central Au+Au collisions. The vertical lines and square brackets show the statistical and systematic errors, respectively.}
\end{figure}

In the coalescence picture, the coalescence parameter, $B_A$, reflects the probability of nucleon coalescence, which is related to the local nucleon density. The reference proton spectra for coalescence parameter extraction are from previous measurements by STAR~\cite{Adamczyk:2017nof}. In the left panel of Fig.~\ref{coal}, the $p_T/A$ dependence of $B_3$ is shown at $\sqrt{s_\mathrm{NN}}=$ 7.7 and 200 GeV for different collision centralities. It is found that $B_3$ increases with increasing $p_T/A$, which might suggest an expanding collision system. The value of $B_3$ decreases from peripheral to central collisions, which can be explained by a decreasing source volume.
In the right panel of Fig.~\ref{coal}, we compare the results of $B_2$ and $\sqrt{B_3}$ in the 0-10\% collision centrality at $p_T/A=$0.65 GeV/c. It is found that $B_2$ and $\sqrt{B_3}$ are consistent within uncertainties except for $\sqrt{s_\mathrm{NN}}=$ 200 GeV. Below 20 GeV, the coalescence parameters $B_2$ and $\sqrt{B_3}$ decrease with collision energy implied that particle-emitting source increase. When $\sqrt{s_\mathrm{NN}}$ $>$ 20 GeV, the decreasing trend seems to change and saturate at BES energy, which might imply a dramatic change of the equation of states of the medium in the collision. At $\sqrt{s_\mathrm{NN}}=$ 200 GeV, $B_2$ and $\sqrt{B_3}$ are different, which might due to the production mechanism for $d$ and $t$ are different.

\subsection{Integral yields and particle ratios}
\begin{figure}[!htp]
\centering
\centerline{\includegraphics[width=1.\textwidth]{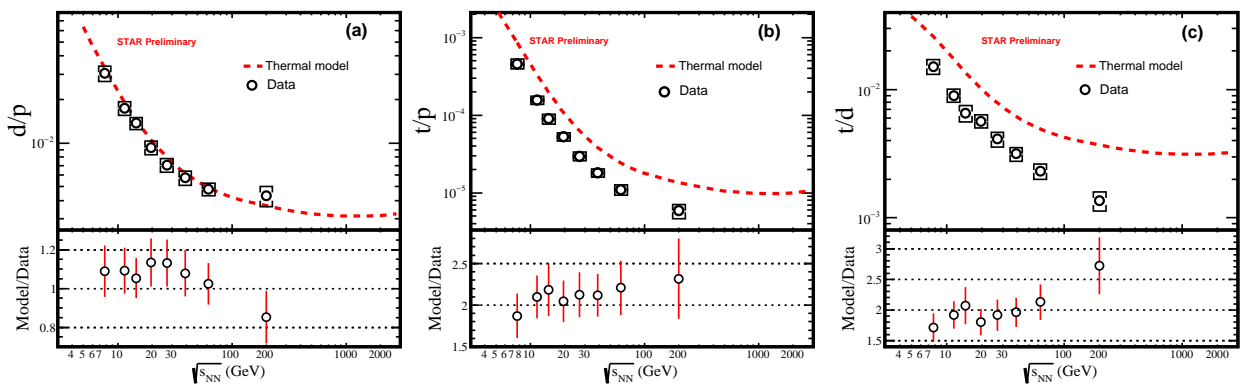}}
\caption{\label{ratios} $d/p$ (left panel), $t/p$ (middle panel), and $t/d$ (right panel) ratios for 0-10\% central Au+Au collisions at BES-I energies. The dotted lines are thermal model predictions. The errors combined the systematic and statistical uncertainties.}
\end{figure}

The yields of triton and deuteron are obtained by measured $p_T$ range and extrapolated to the unmeasured $p_T$ regions with various parameterizations. The extrapolation is done by the individual blast-wave fits~\cite{Schnedermann:1993ws}. We also show the yield ratios of $d/p$, $t/p$, and $t/d$ as a function of collisions energy in 0-10\% central Au+Au collision in Fig.~\ref{ratios}. The proton yields are corrected for the feed-down contribution~\cite{Adamczyk:2017nof}. The dashed lines are the thermal model calculations, which employ parameters established from the analysis of light hadron production in relativistic nuclear collisions. The thermal model describes the ratios of $d/p$ very well but overestimate the $t/p$ and $t/d$ particle ratios~\cite{Andronic:2010qu, Yu:2018ijt}. 
\subsection{Neutron density fluctuation}
\begin{figure}[tbh]
\centering
\includegraphics[width=0.7\textwidth]{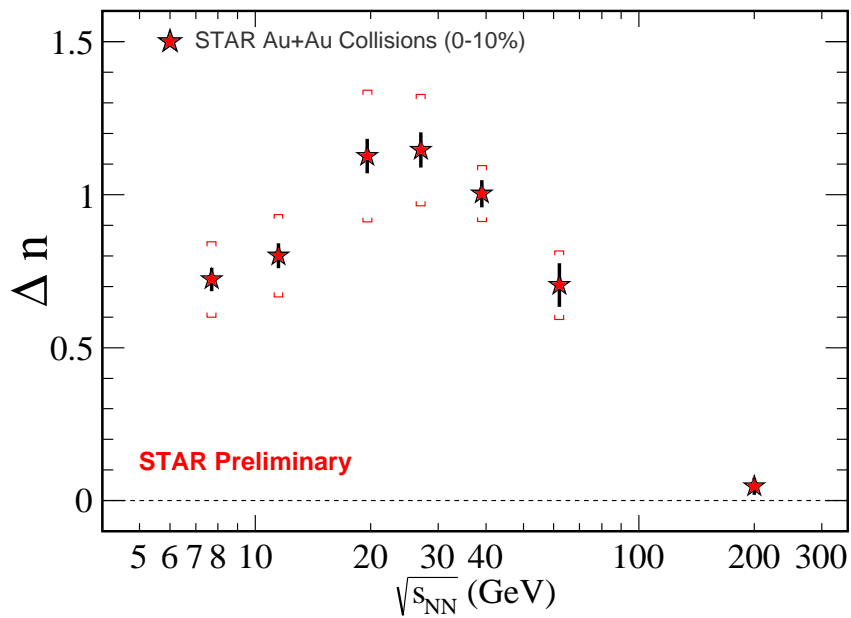}
\caption{\label{deltan}The neutron density fluctuation, $\Delta$n, shows a non-monotonic behavior on collision energy and peaks at around 20 GeV. The vertical lines and square brackets show the statistical and systematic errors, respectively.}
\end{figure}

The neutron density fluctuation, $\Delta n$, extracted from yield of light nuclei is presented in Fig.~\ref{deltan}. A maximum around $\sqrt{s_\mathrm{NN}}=$ 20 GeV is found. This non-monotonic behavior indicates that the density fluctuations become the largest in collisions at this energy. When the evolution trajectory approaches the critical endpoint, the correlation length increases dramatically~\cite{Berdnikov:1999ph}, and the density fluctuation enhances accordingly and reaches its maximum. The baryon density fluctuations are expected to be negligible if the phase transition from QGP to the hadronic matter is a crossover. Therefore, nucleon density fluctuations at kinetic freeze-out in relativistic heavy-ion collisions may provide a unique probe to the critical endpoint in the QCD phase diagram, which need further studies, especially the precise measurements and theoretical understanding.
\section{Conclusions}
We present systematic studies of $d$ and $t$ production in heavy-ion collisions at $\sqrt{s_\mathrm{NN}}=$7.7-200 GeV. The extracted coalescence parameter $B_3$ decreases with increasing collision energy and increases with increasing transverse momentum. The values of $B_{2}$ for deuteron and $B_{3}$ for triton decrease as collision energy increases and seem to reach a minimum around 20 GeV, indicating a change in the equation of state. The values of $B_{2}$ and $\sqrt{B_3}$ from central collision shows the same trends excepted at 200 GeV, which might imply a different formation mechanism for triton and deuteron at the top RHIC energy. The $d/p$ ratio can be reproduced by the thermal model but it cannot describe the triton production. We also measured the collision-energy dependence of neutron density fluctuation and a non-monotonic energy dependence with a peak at around 20 GeV was found, which may indicate that the thermodynamic evolution trajectories of the system pass through the critical region.  

\section*{Acknowledgement}
This work is supported by the National Natural Science Foundation of China under Grants 
(No. 11890711, 11575069, 11828501 and 11861131009), Fundamental Research Funds for the Central Universities No. CCNU19QN054.

\end{document}